\documentclass[11pt]{article}
\usepackage{aaspp4}
\newcommand{\sgra}{\mbox{Sgr~A${}^*$}}
\newcommand\clean{\textsc{clean}}
\newcommand\aips{\textsc{aips}}
\newcommand{\ths}{\theta_s}
\newcommand{\thi}{\theta_i}
\newcommand{\delne}{\delta n_e}

\newcommand{\thxgal}{\theta_{\mathrm{xgal}}}
\newcommand{\thgal}{\theta_{\mathrm{Gal}}}
\newcommand{\delgc}{\Delta_{\mathrm{GC}}}
\newcommand{\dgc}{D_{\mathrm{GC}}}
\newcommand{\mjybm}{\mbox{mJy~beam${}^{-1}$}}
\newcommand{\hoh}{H${}_2$O}
\lefthead{Lazio \& Cordes}
\righthead{Galactic Center Survey for Extragalactic Sources}

\received{1997 July~23}
\paperid{36896}

\begin{document}

\title{Hyperstrong Radio-Wave Scattering in the Galactic Center. I.\\
	A Survey for Extragalactic Sources Seen Through \\
	the Galactic Center}

\author{T.~Joseph~W.~Lazio\altaffilmark{1}}
\affil{Naval Research Laboratory, Code~7210, Washington, DC  20375-5351;
	lazio@rsd.nrl.navy.mil}

\and 

\author{James~M.~Cordes}
\affil{Department of Astronomy and National Astronomy \& Ionosphere
	Center, \\
	Cornell University, Ithaca, NY  14853-6801; \\
	cordes@spacenet.tn.cornell.edu}

\authoraddr{T. Joseph W. Lazio
            NRL, Code 7210
            Washington, DC  20375-5351}
\altaffiltext{1}{NRC-NRL Research Associate}

\begin{abstract}
The scattering diameters of \sgra\ and several nearby OH masers
($\approx 1\arcsec$ at 1~GHz) indicate that a region of enhanced
scattering is along the line of sight to the Galactic center.  The
scattering diameter of an extragalactic source seen through this
scattering region will be larger by the ratio of the Sun-GC distance
to the GC-scattering region separation.  This ratio could be a factor
of a few, if the scattering region is far from the GC and only a
random superposition with it, to more than 100, if the scattering
region is within the \hbox{GC}.  We have used the VLA to survey 10
(11) fields at 20~cm (6~cm) that are between 7\arcmin\ and 137\arcmin\
from \sgra.  Our objective was to identify extragalactic sources and
measure their scattering diameters so as to constrain the
GC-scattering region separation.  In order to find sources within
these fields, we have employed pdf\clean, a source detection algorithm
in which sources are identified in an image by comparing the intensity
histogram of the image to that expected from a noise-only image.  We
found over 100 sources, with the faintest sources being approximately
3~mJy.  The average number of sources per field is approximately 10,
though fields close to \sgra\ tend to contain fewer sources.  In a
companion paper we combine our survey with previous observations of
the GC, and we assess the likelihood that the scattering region is so
close to the GC that the resulting scattering diameters cause
extragalactic sources to be resolved out by our observations.

A number of Galactic sources is included in our source catalog.  We
discuss the double-lobed source 1LC~359.872$+$0.178, potentially an
X-ray quiet version of 1E~1740.7$-$2942, a shell-like structure with a
central point source, and a possible radio transient.

\end{abstract}


\section{Introduction}\label{sec:gc.intro}

If viewed through a plasma containing density fluctuations, an
otherwise unresolved source will have a visibility, as measured by an
interferometer of baseline length~$b$, of
\begin{equation}
V(b) = \exp\left[-\frac{1}{2}D_\phi(b)\right].
\label{eqn:visibility}
\end{equation}
The phase structure function, $D_\phi(b) \equiv
\langle[\phi(0)-\phi(b)]^2\rangle$, is a measure of the phase
perturbations, on a length scale~$b$, imposed on a propagating
electromagnetic wave by fluctuations in the electron density.  

For a plane wave impinging on this scattering region
\begin{equation}
D_\phi(b) 
 = 8\pi^2r_{\mathrm{e}}^2\lambda^2\int_0^D dz\,\int_0^\infty dq\,q[1 - J_0(bq)]P_{\delne}(q, z),
\label{eqn:structurefunction1}
\end{equation}
where $r_{\mathrm{e}}$ is the classical electron radius, $J_0(x)$ is
the zeroth-order Bessel function, $P_{\delne}$ is the spatial spectrum
of the density fluctuations, and the integral over $z$ is taken
\emph{from the source to the observer}.  If the source of radiation is
close to or embedded within the scattering medium, so that the medium
is illuminated by spherical wavefronts, the argument of the Bessel
function is $bq(z/D)$ (\cite{i78}); the factor~$z/D$ accounts for the
divergence of spherical waves.

The apparent angular diameter of the source is determined by the width
of the visibility function, and, hence, by how quickly $D_\phi(b)$
decreases as a function of~$b$.  Since $z/D < 1$, the difference in
the form of the phase structure function for plane and spherical
wavefronts means that sources close to the medium will show smaller
angular diameters than those far from it.  Hence, by comparing the
scattering diameters of Galactic and extragalactic sources along
similar lines of sight, one can constrain the \emph{radial} location
of the scattering material.

Toward the Galactic center, the observed diameter of \sgra\ scales as
$\lambda^2$ over the wavelength range 30~cm to~3~mm (Davies, Walsh, \&
Booth~1976; \cite{rogersetal94}), as expected if very strong
interstellar scattering from microstructure in the electron density
determines the observed diameter.  Maser spots in OH/IR stars within
25\arcmin\ of \sgra\ also show enhanced angular broadening
(\cite{vfcd92}; \cite{fdcv94}).  The scattering disks of \sgra\ and
many of the OH masers are observed to be anisotropic as well
(\cite{vfcd92}; \cite{bzkrml93}; \cite{krichbaumetal93};
\cite{fdcv94}; \cite{y-zcwmr94}); in the case of \sgra, its scattering
disk is anisotropic at least over the wavelength range 21~cm to 7~mm.
These observations indicate that a region of enhanced scattering with
an angular extent of at least 25\arcmin\ (60~pc at 8.5~kpc) is along
the line of sight to \sgra.  At 1~GHz the level of angular broadening
produced by this scattering region is roughly 10 times greater than
that predicted by a recent model for the distribution of free
electrons in the Galaxy (Taylor \& Cordes~1993, hereinafter
\cite{tc93}), even though this model includes a general enhancement of
scattering toward the inner Galaxy.

Because all of the sources observed through this region have thus far
been Galactic sources, with (presumably) approximately the same
location (i.e., in the Galactic center), the radial location of the
scattering region is unconstrained.  The scattering region could be
local to the Galactic center, within approximately 100~pc from the
Galactic center---which we refer to as the GC model---or the region
could be a random superposition and more than 1~kpc from the
GC---which we refer to as the RS model.  In the GC model, the region
would be a site of excess scattering, and presumably arises from
processes unique to the GC; in the RS model, the level of scattering
in the region would be high, but not unusually so.

Previous estimates for the location of the scattering region have
ranged from 10~pc to 3~kpc.  Ozernoi \& Shisov~(1977) concluded that
an ``unrealistic'' level of turbulence is implied unless the region is
within 10~pc of the \hbox{GC}.  The level of turbulence they
considered unrealistic, however, namely $\sqrt{\langle
n_{\mathrm{e}}^2\rangle}/\langle n_{\mathrm{e}}\rangle \sim 1$, does
appear to occur elsewhere in the interstellar medium (\cite{s91}).
Further, van~Langevelde et al.~(1992) showed that the free-free
absorption toward \sgra\ would be excessive unless the scattering
region was at least 0.85~kpc from the GC, though suitable adjustment
of free parameters (outer scale and electron temperature) can decrease
the limit to 0.03~kpc.  With the free-free absorption they also placed
an upper limit on the region's distance from the GC of 3~kpc.
Although the GC model is attractive for phenomenological reasons,
other sites of enhanced interstellar scattering are found throughout
the Galaxy (e.g., NGC~6634, \cite{mrgb90}; Cyg~X-3, \cite{mmrj95}) and
the mean free path for encountering such a region is approximately
8~kpc (\cite{cwfsr91}).

Identifying the location of the scattering region may provide clues to
the origin of the scattering.  The density fluctuations responsible
for interstellar scattering are believed to be generated by velocity
or magnetic field fluctuations (\cite{h84}, 1986; Montgomery, Brown,
\& Matthaeus~1987; \cite{s91}; \cite{sg94}; \cite{gs95}).  Velocity or
magnetic field fluctuations are also a natural means for inducing
anisotropy in the density fluctuations and thereby in the scattering
disks.  If this supposition is correct, the amplitude of the density
fluctuations may provide a measure of the coupling between the density
and velocity or magnetic field fluctuations or, more generally,
provide information about the small-scale velocity or magnetic field
in the scattering region.  However, because the radial location of the
scattering region is unconstrained, relevant quantities, e.g., the rms
density, are uncertain by a factor of $\delgc/\dgc$, where $\delgc$ is
the GC-scattering region separation and $\dgc$ is the GC-Sun distance.

Observations of extragalactic sources viewed through the scattering
region could constrain $\delgc$; however, few extragalactic sources
have been identified toward the \hbox{GC}.  The two sources closest to
\sgra\ are B1739$-$298 (\cite{dkvgh83}) and GPSR~0.539$+$0.263
(Bartel~1994, private communication), which are 48\arcmin\ and
40\arcmin\ from \sgra, respectively.  Neither of these is within the
region of enhanced scattering defined by the OH masers.

This paper reports VLA and VLBA observations of potential
extragalactic sources seen through the \hbox{GC}.
Section~\ref{sec:gc.observe} describes the observations and data
reduction, Section~\ref{sec:catalog} discusses the identification of
potential extragalactic sources and presents the catalog of sources,
Section~\ref{sec:sources} discusses certain Galactic sources found in
our VLA survey, and Section~\ref{sec:gc.conclude} discusses our
results and presents our conclusions.  A companion paper (Lazio \&
Cordes~1998, hereinafter \cite{lc98}) combines the results of this
paper with the previous observations of OH and \hoh\ masers and
free-free emission in a likelihood analysis that constrains the
angular extent and radial location of the scattering region.
\cite{lc98} also discusses the physical conditions inside the
scattering region.

\section{Observational Program}\label{sec:gc.observe}

The scattering diameter of a compact extragalactic source viewed
through the scattering region toward the GC is (\cite{vfcd92})
\begin{equation}
\thxgal = \frac{\dgc}{\delgc}\thgal,
\label{eqn:xgalsize}
\end{equation}
where $\thgal$ is the characteristic diameter of a GC source.
Throughout this paper we will adhere to the convention that scattering
diameters are the full width at half maximum of the source's intensity
distribution.  We adopt $\dgc = 8.5$~kpc for the GC-Sun distance (at
this distance $1\arcmin = 2.5$~pc), and the observed diameter of
\sgra\ at 1~GHz is 1\farcs3.  Figure~\ref{fig:xgalsize} shows
$\thxgal$ as a function of $\delgc$.  If the RS model is correct and
$\delgc \gtrsim 1$~kpc, extragalactic source diameters should be a few
arcseconds; if the GC model is correct and $\delgc \approx 100$~pc,
source diameters could exceed 1~\emph{arc\,min}.

Motivated by the prediction illustrated in Figure~\ref{fig:xgalsize}, we
undertook a program to identify extragalactic sources toward the GC,
measure their angular diameters, and, thus, constrain $\delgc$.
Scattering diameters of this magnitude are the province of the VLA,
and Section~\ref{sec:vlaobserve} describes the source selection
criteria and the VLA observations in our survey.  We also conducted
VLBI observations on a subset of the sources detected in our VLA
survey; Section~\ref{sec:gc.vlbaobserve} describes these observations.

\subsection{VLA Observations}\label{sec:vlaobserve}

Our initial source list was compiled from previous observations of the
GC and contained 15 sources, judged likely to be extragalactic either
from spectral information or because the sources appeared compact.
The fields observed are listed in Table~\ref{tab:point} and are shown
in Figure~\ref{fig:point}; some fields contain more than one source.
The fields range in distance from~7\arcmin\ to~137\arcmin\ from \sgra.
Such a large range in angular distances from \sgra\ serves two
purposes.  First, some of the fields could show heavy scattering, but
not at the level seen toward \sgra.  These fields will serve as
controls.  Second, the scattering region could be patchy so that the
sources B1739$-$298 and GPSR~0.539$+$0.263 are viewed through
``holes'' in the intense scattering region,

The observations were conducted on 1994 May~15 and~16 in two,
eight-hour sessions with the VLA in the BnA configuration.  We
observed at both 6 and~20~cm.  At 20~cm, the array's resolution is
5\arcsec\ and structure as large as approximately 1\arcmin\ can be
detected.  It is therefore well-matched to a wide range of scattering
diameters, and we hoped to identify extragalactic sources and detect the
characteristic $\lambda^2$ dependence of angular broadening for their
angular diameters.  For this reason, the AC and BD IFs were centered on
1281 and~1658~MHz, respectively, the widest possible separation
allowed by the VLA hardware and radio frequency interference (RFI)
environment.  We planned to use the 6~cm observations to provide
spectral information to assist in the identification of extragalactic
sources, and, possibly, to determine intrinsic structure and diameters for
sources.  The IFs for these observations were centered at 4823
and~4873~MHz.  Anticipating the need for a wide field of view to allow
for the subtraction of any confusing sources, the observations were
conducted in a pseudo-continuum mode.  Both right and left circular
polarization were recorded in seven, 3-MHz channels; because of edge
effects, the total usable bandwidth is 18~MHz per \hbox{IF}.  At both bands
the resulting bandwidth smearing is negligible even at 30\arcmin\ from
the pointing center.

During each eight-hour session, each source was observed at two hour
angles for each band.  The observations were conducted in snapshot
mode, with a duration of 5~min.\ per hour angle per field.  The total
observing time per field was 20~min.\ at~20~cm, obtained over 4 hour
angles, and 10~min.\ at~6~cm, obtained over two hour angles.  The
resulting theoretical noise level in our images is 0.4~\mjybm\ at both
bands.

The data were edited and calibrated in the standard fashion within
\aips.  The flux scale was set with observations of 3C286 and 3C48;
frequent observations of B1748$-$253 were used to calibrate the
visibility phases.  For some of the fields at~20~cm, which showed
possible gain errors in the dirty images, we also self-calibrated
before undertaking the mapping and {\clean}ing described in
\S\ref{sec:catalog}.  Typically one iteration of phase-only
self-calibration was applied to the data.  The resulting reduction in
the off-source rms intensity in images was a factor of 1.5 to~2.

\subsection{VLBA Observations}\label{sec:gc.vlbaobserve}

As we discuss later, we find a number of compact sources at~20~cm that
do not display a $\lambda^2$ dependence for the angular diameter.
Compact sources not showing scattering could arise from several
possibilities: First, Galactic sources and extragalactic sources not
behind the enhanced scattering region could have diameters much
smaller than that of \sgra---the TC93 model predicts a diameter of
0\farcs1 at 1~GHz.  Second, if the scattering region is far from
\sgra, extragalactic source diameters should be approximately double
that of \sgra.  That is, source diameters would be 1\farcs6
at~1281~MHz and 0\farcs95 at~1658~MHz.  These diameters are
sufficiently smaller than the typical beam at~20~cm ($\approx
5\arcsec$) that the ``deconvolution'' of the beam from the measured
angular diameters could obscure the effect of scattering, particularly
in the presence of noise and at~1658~MHz.  Lastly, the scattering
screen could be patchy.

In order to test the possibility that the diameters of these compact
sources are similar to those predicted by the RS model, we undertook a
program of VLBI angular broadening measurements on a small subset of
compact sources detected in the VLA program.  These sources are listed
in Table~\ref{tab:vlbisources} and are shown in
Figure~\ref{fig:point}.

The sources were observed with a subset of the VLBA---stations PT, LA,
KP, FD, OV, and~NL---and the phased \hbox{VLA}.  Observation
wavelengths were 1.3, 3.6, and~6~cm.  The wavelengths of observation
and VLBI stations were chosen so that the resolution of the array was
well-matched to the diameters expected for sources showing broadening
typical of the RS model.  The resolutions of this array are 1.2~mas at
1.3~cm, 3.9~mas at 3.6~cm, and 6.5~mas at 6~cm; the scattering disk of
\sgra\ is 2~mas at 1.3~cm, 17~mas at 3.6~cm, and 52~mas at 6~cm
(\cite{lbkrzgm93}; \cite{y-zcwmr94}).

The observations were conducted on 1995 September~22 and~23.  Dual
polarization was recorded at all three wavelengths, in four 8-MHz IFs,
for a total bandwidth of 32~MHz.  Data at two hour angles were
obtained at 1.3 and~3.6~cm and one hour angle was obtained at 6~cm.
However, fringes were not found---even for NRAO~512, an approximately
1~Jy source observed to assist with the fringe finding---during the
first hour angle at 1.3 and~3.6~cm.  The lack of fringes is due, in
part, to receiver problems, as indicated by station logs.  The
resulting total time on source for all three wavelengths is 5~min.,
obtained at one hour angle.

The data were analyzed within \aips.  Editing was performed using
station-supplied logs; some additional editing, almost exclusively
near scan boundaries, was performed later.  Amplitude calibration for
the VLBA stations was performed using $T_{\mathrm{sys}}$ measurements
for those stations.  For baselines including the phased VLA, a source
flux density is required.  Two of the program sources, B1739$-$298 and
1LC~359.872$+$0.178, have fluxes measured over most of the wavelength
range of interest (\cite{zhbwp90}; Anantharamaiah \& Goss~1994,
private communication; \S\ref{sec:catalog}).  The fluxes for the other
two program sources were extrapolated from lower frequency
measurements assuming the sources had flat spectra.  Fluxes for
B1741$-$312 were taken from the VLA calibrator manual (\cite{t97}) and
for \sgra\ from Lo et al.~(1993).  An opacity correction, based on
station-supplied meteorological data, was applied to the 1.3~cm data.
Fringe fitting was performed in two steps.  First, the delays across
individual IFs due to the electronics within the IF were determined.
Then the rate, delay between IFs, and the residual delay across
individual IFs were determined.  Finally, the data were corrected for
the shapes (amplitude and phase) of the IF bandpasses at each station.
Further analysis on these VLBI data were conducted after integrating
them over the full 32~MHz bandpass and for 30~s in time.

\section{Source Detection, Identification, and Angular Diameter
	Determination}\label{sec:catalog}

This section describes the process of source detection in the VLA
fields, discusses the extraction of angular diameters from the VLBI
observations, and presents the resulting catalogs and identifications
of these sources.  Our initial list of candidate extragalactic sources
contained 15~targets.  Because of the large field of view afforded by
the pseudo-continuum observations, we were able to identify well over
100~objects.

\subsection{Source Detection in the VLA Fields}\label{sec:sourcedetect}

The intensity distribution of a noise-only image from a multiplying
correlator is a gaussian (\cite{f88}).  As Zepka, Cordes, \&
Wassermann~(1994) demonstrated, by comparing the shape of the expected
noise-only distribution to that determined from the image itself, it
is possible to detect sources below a nominal signal-to-noise
threshold.  We adapted this source detection procedure in the
following manner:

\begin{enumerate}
\item Map the primary beam out to the 15~dB level ($\approx 30\arcmin$
at~20~cm and $\approx 7\arcmin$ at~6~cm) and form the intensity
histogram of this image;

\item Compare this histogram to a gaussian having the same mean and
variance as the image;

\item If the intensity histogram shows deviations, at positive
intensities, from a gaussian, locate the deviant pixels in the image,
then map and \clean\ a small region around them; and

\item Subtract the \clean\ components of these sources from the
visibility data.
\end{enumerate}
This procedure was iterated until the intensity histogram showed good
agreement with a gaussian or until deviant pixels could not be
identified as sources.  Some fields showed symmetrical intensity histograms,
though with non-gaussian wings.  Such non-gaussian, symmetrical wings
can occur because of sidelobes from distant sources or because of
\hbox{RFI}.

Our procedure is iterative, in contrast to that described by Zepka et
al.~(1994) who applied it to X-ray images, because we make use of
aperture synthesis images.  The beam of an aperture synthesis
instrument has high sidelobes (reaching $\approx 10$\% of the main
beam peak), and the sidelobes extend over the entire image.  If not
processed in an iterative manner, sidelobes from stronger sources can
obscure fainter sources or be identified mistakenly as sources
themselves.  Isaacman~(1981) used a similar iterative procedure in
searching for planetary nebulae in the GC, though he used a local
signal-to-noise ratio threshold to identify sources.

We call this procedure pdf\clean\ (\cite{lc96}; Cordes, Lazio, \&
Sagan~1997): Traditional implementations of \clean\ attempt to reduce
an image to the noise level (and then produce a final image by adding
\clean\ components to this residual image and smoothing).  Moreover, in
Clark's~(1980) version of \clean, deconvolution of a beam from an
image is split into ``major'' and ``minor'' cycles.  In the major
cycle image pixels from which the beam will be deconvolved are
identified using a histogram.  We view our method of reducing an image
histogram to a noise-only histogram, and in the process identifying
image pixels constituting potential sources, as a natural extension of
\clean\ methods in general and of Clark's~(1980) method in particular.

The mapping and {\clean}ing of sources was done independently in each
frequency channel.  In deciding whether or not deviant pixels
constituted a source, we required that sources appear in all six
channels and that the {\clean}ed flux be positive.  After mapping and
{\clean}ing a source, a primary beam correction was applied (\aips\
task \texttt{PBCOR}) and the intensity was averaged over the channels.
At~6~cm the images were also averaged over the two IFs.

We used this procedure on all fields at~6 and~20~cm.  For fields
within approximately 1\arcdeg\ of \sgra, we made minor modifications
to the mapping and {\clean}ing procedure.  These fields are distinguished
by their intense, extended emission.  The emission is sufficiently
intense that it is not filtered out completely by the shortest
baselines in the BnA configuration, but neither is it sampled
sufficiently to be reconstructed accurately by BnA observations alone.
We therefore imposed inner $u$-$v$ limits, of 1--4~k$\lambda$, during
the mapping.  These inner $u$-$v$ limits have the effect of reducing
further our sensitivity to large angular structures.  For the fields
with these inner $u$-$v$ limits, the largest detectable angular scale
is approximately 30\arcsec.

For these fields we also knew \textit{a priori} the approximate area
over which extended emission would be found (\cite{l88}; Liszt~1994,
private communication).  The diameters of these regions are arcminutes, in
contrast to the typical diameters of the regions we {\clean}ed in other
fields, a few to a few tens of arcseconds.  Within these
arcminute-sized regions, we therefore accepted as sources only those
peaks of emission for which \clean\ components were found in the inner
five channels, excluding two channels to allow for edge effects.  Many
of the sources thus identified appear to be portions of extended
emission seen in lower resolution images.

As noted in \S\ref{sec:vlaobserve}, the theoretical noise level in our
images is 0.4~\mjybm.  Through the use of pdf\clean\ we obtained rms
noise levels of 0.5--2~\mjybm\ for the various fields, with the fields
closest to \sgra\ having the higher noise levels.  In \cite{lc98} we
shall need the minimum detectable source flux density,
$S_{\mathrm{min}}$, for each of the survey fields in order to estimate
the expected number of extragalactic sources in a field.  We
determined $S_{\mathrm{min}}$ from each residual image, i.e., an image
formed after subtracting all sources detected in a field.  Only the
central portion of a residual image, for which the primary beam
correction was less than 2\%, was used.  This region is 4\arcmin\ in
size and contains about $3 \times 10^4$ pixels.  The region is large
enough that $1/\sqrt{N}$ is small, but small enough that the primary
beam correction is unimportant.  We took $S_{\mathrm{min}} \approx
4\sigma$; the minimum flux densities are 2--8~mJy.

The sources detected by our methods are cataloged in
Tables~\ref{tab:sources} (20~cm) and~\ref{tab:c-sources} (6~cm).  We
adopt ``1LC'' as the catalog designation for the sources we have
detected.

If a source has been detected at both 1281 and~1658~MHz, we provide
the coordinates from the 1658~MHz identification.  Whenever possible
we have made gaussian fits, in the image plane (\aips\ task
\texttt{JMFIT}), to the sources and report the peak and integrated
intensity and the ``deconvolved'' FWHM source diameter.  The
deconvolution is accomplished by subtracting in quadrature the beam
size ($\approx 5\arcsec$) from the fitted diameter.  If a gaussian
could not be fit to a source, the source was defined to be the
polygonal region within which the intensity exceeded $2\sigma$; the
peak and integrated intensity within this region are tabulated.  The
size reported is the product of the number of beams covering the
source and the beam size.

For sources detected at both 1281 and~1658~MHz, we report the spectral
index ($S_\nu \propto \nu^{-\alpha}$).  Some of the sources have
extreme spectral indices, with $|\alpha| > 4$.  We believe that these
extreme spectral indices probably reflect a combination of the closely
spaced observing frequencies and the primary beam correction rather
than the actual spectra.  Because the magnitude of the primary beam
correction depends upon the distance from the pointing center, we also
give this distance.

Table~\ref{tab:c-sources} for the sources from the 6~cm survey is
similar in format to Table~\ref{tab:sources} for the sources from the
20~cm survey.  We combined the IFs at~6~cm, so we report only one peak
intensity, integrated flux, and source diameter measurement.  Also,
the spectral index is calculated between 6 and~20~cm.

\subsection{Source Identification}\label{sec:actualcatalog}

We have compared our sources to the Columbia Galactic plane survey
(catalog acronym GPSR) of Zoonematkermani et al.~(1990) and Helfand et
al.~(1992).  They surveyed the central Galactic plane, $-10\arcdeg \le
\ell \le 40\arcdeg$ and $|b| \le 1\fdg8$, using the VLA, with a
similar resolution, at~1.4~GHz.  They searched for sources in
15\arcmin\ fields and obtained 5$\sigma$ sensitivities of 5~mJy for
fields with $|b| \approx 1\fdg5$ to 50~mJy for fields near \sgra.
They also compared their catalog to several other surveys and were
able to make a number of matches.  We classify those sources common to
both our survey and the GPSR as Galactic, extragalactic, or unknown.
We consider as Galactic those sources having positional coincidences
with supernova remnants, planetary nebulae, stars, Einstein X-ray
sources, and IRAS sources.  A small number of sources are also
classified as Galactic by comparison with other surveys of the GC
(\cite{y-zm87}; \cite{l88}).  Sources are considered extragalactic if
they were identified with galaxies in the \hbox{GPSR}.  A small number
of sources are also classified as extragalactic by comparison with the
VLA calibrator manual (\cite{t97}), the VLBI observations reported
here (\S\ref{sec:vlbisources}), or from other GC VLBI observations
(Bartel~1994, private communication).

Figure~\ref{fig:srccat} combines both our survey and the GPSR,
indicating whether the sources are Galactic, extragalactic, or as yet
unidentified.

Two aspects of the source distribution are evident immediately.
First, the fields we have observed have a higher density of sources as
compared to those in the GPSR, resulting from both our improved
sensitivity and our use of \hbox{pdf\clean}.  Second, there is a
noticeable paucity of sources near Sgr~A, and particularly to the
northwest.

This paucity could arise from three effects.  First, the increase in
extended emission near Sgr~A will result in an increase in the system
temperature and a decrease in the sensitivity.  Second, the majority
of the Galactic radio sources found in the GPSR are \ion{H}{2} regions
(\cite{bwmhz92}; \cite{hzbw92}).  The inner 100~pc or so of the GC are
not, on average, a site of current, vigorous star formation, which
could contribute to a deficit of Galactic sources (\cite{rg89};
\cite{ms96}).  Finally, if the excess scattering region covers this
portion of the GC and $\delgc$ is sufficiently small, extragalactic
sources could be broadened to the point that they would be resolved
out by both our observations and those of Zoonematkermani et
al.~(1990).  To evaluate these possibilities quantitatively, we
performed a likelihood analysis that is described in \cite{lc98},
where we find that this paucity is due, in part, to hyperstrong
scattering occurring in the GC.

\subsection{Angular Diameters from VLBI
	Measurements}\label{sec:vlbisources} 

For the bandwidth (32~MHz) and integration time (30~s) used, the
expected rms visibility in the total intensity for a VLBA-VLBA
baseline is approximately 10~mJy at~6~cm, 15~mJy at~3.6~cm, and 50~mJy
at~1~cm; the VLBA-VLA baselines should have an rms a factor of
$\sqrt{27}$ smaller than that of the VLBA-VLBA baselines.  We also
considered an integration time of 60~s in an effort to detect sources
or improve the signal-to-noise ratio for weak sources.  Longer
integration times, though allowed by the coherence time at~3.6
and~6~cm, produced few visibilities on any given baseline and were not
used.

If the signal-to-noise ratio in the visibility data for a source is
sufficiently high, we have fit a circular gaussian.  Our limited hour
angle coverage does not justify more complicated models.  For those
sources showing marginal detections, we attempted to assess their
structure using the rms visibility phase, $\sigma_\phi$.  Thompson,
Moran, \& Swenson~(1986) show that, in the low signal-to-noise case,
the visibility modulus is
\begin{equation}
\frac{|V|}{\sigma} 
 \approx \frac{\pi\sqrt{2}}{3}\left(1 - \frac{\sqrt{3}}{\pi}\sigma_\phi\right),
\label{eqn:sigmav}
\end{equation}
where $\sigma^2$ is the variance of the visibility modulus.  If
$|V|/\sigma \to 0$, the distribution of the visibility phase is
uniform on the interval 0 to $2\pi$.  For weak sources we have
compared the visibility phases to a uniform distribution using the
Kolmogorov-Smirnov statistic.  For those baselines for which the
phases are not consistent with a uniform distribution, we compute
$|V|/\sigma$ using equation~(\ref{eqn:sigmav}).

We discuss the results for each source separately.

\subsubsection{\sgra}

We observed \sgra\ primarily as a control source.  It is detected at
all three frequencies.  The source diameters we derive are consistent
with those expected from previous measurements (viz.\
\S\ref{sec:gc.vlbaobserve}): $51.6 \pm 0.8$~mas at 6~cm, $16.6 \pm
0.4$~mas at 3.6~cm, and $2.3 \pm 0.2$~mas at 1.3~cm.

\subsubsection{B1741$\mathit{-}$312 (1LC~357.862$-$0.997)}

This source is a VLA calibrator source (\cite{t97}) and is described
by Backer~(1988) as ``heavily scattered.''  We detect it at all three
frequencies.  While not as heavily scattered as \sgra\ (its scattering
diameter is a factor of 5 smaller), its 1~GHz scattering diameter is
approximately a factor of 2 larger than that predicted by the TC93
model.  Its diameters are $18.2 \pm 0.4$~mas at 6~cm, $7.2 \pm
0.2$~mas at 3.6~cm, and $1.0 \pm 0.1$~mas at 1.3~cm; these diameters
scale as $\lambda^{1.8\pm0.4}$.

\subsubsection{B1739$-$298 (1LC~358.918$+$0.073)}\label{sec:1742-2949}

This source is located approximately 1\arcdeg\ from \sgra.  Most of
its displacement from \sgra\ is in longitude and it provides a
constraint on the extent of the screen to negative longitudes.

At 6~cm this source is detected clearly on short baselines,
Figure~\ref{fig:j1742-2949}, and it has a diameter of $27.9 \pm
1.9$~mas.  At~3.6~cm this source is only marginally detected on the
inner baselines which include the phased VLA; we can place only an
upper limit of 55~mas on its diameter.  At~1~cm this source is not
detected.  This non-detection is consistent with the source's
spectrum: At 3.6~cm the source has a flux of approximately 30~mJy
(\cite{zhbwp90}) and between 6 and~3.6~cm, its spectral index is 1.9
($S \propto \nu^{-\alpha}$).  At 1.3~cm, its expected flux is
therefore approximately 5~mJy, well below our detection threshold.

We have attempted to determine an angular diameter for the source at
18~cm from our VLA observations.  From a super-uniformly--weighted
image we find a diameter of 0\farcs86.  This diameter is likely to be
an upper limit.  Taking $\theta \propto \lambda^\beta$ between 6
and~18~cm, $\beta = 3$, far steeper than predicted even by scattering.

The source's 6~cm diameter is larger than predicted by the TC93 model
(a factor of 5), but smaller than that of \sgra\ (a factor of 2).
Scaling its 6~cm diameter to 18~cm, assuming a $\lambda^2$ dependence
for the angular diameter, its predicted scattering disk at 18~cm is
250~mas.  This diameter is comparable to many of the OH masers with
$\ell < 0\arcdeg$ and $|b| \gtrsim 1\fdg5$ (\cite{vfcd92}), yet $b =
0\fdg07$ for B1739$-$298.  In both the GC and RS models presented in
\S\ref{sec:gc.intro}, extragalactic sources seen through the enhanced
scattering region must have diameters larger than that of \sgra, viz.\
Figure~\ref{fig:xgalsize}.  Galactic sources can have a range of
source diameters, depending upon whether they are behind the
scattering region or not and, if behind the scattering region, their
distance from the region.

This source is extragalactic: A lower limit to its brightness
temperature is $10^6$~K and it is detected in the Texas survey at
365~MHz (\cite{dbbtw96}).  The \ion{H}{1} absorption towards this
source places a lower limit of 25~kpc on its distance
(\cite{dkvgh83}).  The \ion{H}{1} spectra of this source and the VLA
calibrator B1741$-$312 are similar in shape and cover nearly the same
velocity range ($-80$~km~s${}^{-1}$ to 50~km~s${}^{-1}$, with
B1739$-$298 showing additional weak emission to 100~km~s${}^{-1}$).
Further, compact Galactic sources are typically also X-ray sources,
e.g., Cyg~X-1 and the central component of 1E~1740.7$-$2947; there are
no X-ray sources within 5\arcmin\ of this source.

Because this source is extragalactic and its diameter is smaller than
that of \sgra, we cannot use it to constrain the radial location of
the screen.  However, in conjunction with the heavily scattered maser
OH~359.517$+$0.001, we shall use it to constrain the angular extent of
the scattering region to negative longitudes to be less than 1\arcdeg\
(\cite{lc98}).

\subsubsection{1LC~358.439$-$0.211}

We do not detect this source at any of the three frequencies.  The
source's 6~cm flux is 38~mJy (\cite{bwhz94}); there are no higher
frequency observations of this source.  At 6~cm this source should be
a 3--4$\sigma$ detection on VLBA-VLBA baselines.  Several
possibilities exist for the source's non-detection:
\begin{description}
\item[Resolved out:] Table~\ref{tab:vlbisources} lists the angular
diameter the source would need to have in order to be resolved out by
the shortest baseline, PT-\hbox{VLA}.  Becker et al.~(1994) determine
a diameter of 2\farcs2 from a naturally-weighted image; we did not
observe this source in our 6~cm observations. The actual 6~cm diameter
could be much smaller because, in a super-uniformly--weighted image
from our 1658~MHz VLA observations, the source has a diameter of
1\arcsec, only slightly larger than the diameter of B1739$-$298.

\item[Structure:]  Complex structure produces nulls in the visibility
plane.  For example, a double source produces a cosinusoidal visibility
function (\cite{p95}).  If this source is a double with a separation of
200--250~mas, the correlated flux at 6~cm on the innermost baselines
would be reduced substantially below the nominal 38~mJy.

\item[Spectrum:] The source has a spectrum $S \propto \nu^{-0.3}$ between
1.4 and~5~GHz.  If we extrapolate this spectrum to 1.3~cm, the
expected flux is 25~mJy, which would be detectable on only the
VLBA-VLA baselines.  The expected 3.6~cm flux is 33~mJy, which would
be marginally detectable on VLBA-VLBA baselines (via
eqn.~[\ref{eqn:sigmav}]).  If the spectrum steepens at higher
frequencies, these flux densities would be correspondingly reduced.

\item[Variability:] There has been only one 6~cm observation of this
source and two measurements at~20~cm of its flux.  The two
measurements at~20~cm (the GPSR and ours) agree to better than 5\% and
are separated by 5~yr.

\item[Position Error:] Position errors lead to fringe rates and delays.
If the rate and delay windows are too small during the fringe fitting,
fringes will not be found.  The position for the source was obtained from
our VLA observations and is accurate to approximately 1\arcsec.  The
rate and delay windows we used allow position errors of several arcseconds.
\end{description}

Of these possible causes, we favor the spectrum as the reason for the
non-detection at 1.3~cm and source structure for its non-detection at
3.6 and~6~cm.  

\subsubsection{1LC~359.872$+$0.178}\label{sec:gc.sourceJ}

Located approximately 15\arcmin\ from \sgra, displaced largely in
latitude, detection of scattering from this source would place severe
constraints on the angular extent of the screen, and possibly on the
radial distance as well.  Section~\ref{sec:sources} summarizes a
number of measurements of this source; here the focus is on just the
VLBI measurements.

We do not detect this source at any of the three frequencies.  Based
on the flux measurements summarized in \S\ref{sec:sources}, the lack
of detection at~1~cm is consistent with the source's spectrum.
At~3.6~cm the flux density is comparable to the expected rms
visibility, and, at best, only a marginal detection using the
visibility phases could be obtained (eqn.~\ref{eqn:sigmav}).  At~6~cm
our attempts to detect this source have included both global and
baseline-based approaches to fringe fitting, and, in the global fringe
fitting, utilizing different rate and delay windows and different
solution intervals.

The lack of a detection at~6~cm is most likely due to source being
sufficiently extended as to be resolved out.  The shortest baselines
in our VLBI array are those on the LA-PT-VLA triangle which are
sensitive to angular structure on a scale of approximately 0\farcs2.
This source is a double, with a component separation of 4\arcsec.  The
stronger component, the one observed, has a flux of~20~mJy at~6~cm.
Yusef-Zadeh, Cotton, \& Reynolds~(1998) have presented higher
resolution VLA observations of this source at~2 and~6~cm.  At~6~cm
they find the stronger component to have a diameter of approximately
$0\farcs4$.  Subsequent to the VLA observations we present here, we
obtained slightly higher resolution observations at~6~cm; we find a
diameter similar to that found by Yusef-Zadeh et al.~(1998).  Further,
they show that at~2~cm, the structure of the stronger component is
resolved into two slightly extended subcomponents with a
``barrel-shaped appearance.''  The other component is unresolved
(\cite{y-zcr98}), but is sufficiently weak ($\approx 5$~mJy) that we
would not have been able to detect it.

\subsubsection{1LC~0.846$+$1.173}

This source is not detected at any of the three frequencies.  As for
1LC~358.439$-$0.211, we quote the angular diameter the source would need
in order to be resolved out.

Of the causes listed for the non-detection of 1LC~358.439$-$0.211, the
spectrum is the most likely cause for this source's non-detection at
1.3 and~3.6~cm with source structure the cause at 6~cm.  There is no
measurement of this source's flux above 1.4~GHz: It was sufficiently
far from the center of the closest field at~6~cm that we have no
observations of it and it was outside the Galactic latitude limit of
the survey by Becker et al.~(1994).  It is detected in the Texas
survey (\cite{dbbtw96}) and has a 0.3--1.4~GHz spectral index of 0.9.
At 6~cm we expect a flux of 50~mJy, comparable to B1739$-$298.
Variability is an unlikely cause as the two flux determinations
at~20~cm (the GPSR and ours) agree to better than 5\%.

We observed this source, even though it has a steep spectrum, because
it is at fairly high latitude.  Most of our fields and most of the
OH/IR stars with measured angular diameters are distributed in
longitude, so that there are few constraints on the latitude
distribution of scattering.  A measured angular diameter could have
improved our knowledge of the latitude distribution of scattering.  As
it happens, this source is drawn from a field in which the number of
sources is consistent with that expected from extragalactic source
counts (\cite{lc98}), implying that scattering is unlikely to dominate
this source's angular diameter and that it is unlikely to be resolved
out.

\section{Galactic Sources from the Survey}\label{sec:sources}

Our primary aim in this project was to identify extragalactic sources
suitable for angular broadening measurements.  In the course of our
survey, we found a number of Galactic sources, a subset of which is
described here.

\subsection{1LC~359.872$+$0.178}

The source 1LC~359.872$+$0.178 (source 35W44, \cite{i81}; source~J,
\cite{y-zm87}; GPSR~359.873$+$0.179, \cite{zhbwp90}) is located
approximately 15\arcmin\ from \sgra\ and potentially within the region
of enhanced scattering detected by van~Langevelde et al.~(1992) and
Frail et al.~(1994).  At high frequencies it has two components,
suggestive of an extragalactic source.  If it is extragalactic, it
sets important constraints on the scattering region in front of \sgra.
Namely, it implies either that the RS model is correct, or that, if
the GC model is correct, the scattering region is patchy or does not
extend to more than 15\arcmin\ in latitude.  Alternately, the source
could be Galactic, in which case it could be an X-ray--quiet version
of 1E~1740.7$-$2942 (\cite{mrcpl92}).  Yusef-Zadeh et al.~(1998) argue
that the two components are unrelated---based on their slightly
different spectral indices, and that the stronger of the two is a
candidate young supernova remnant.

This section summarizes observations of this source at a variety of
wavelengths.  We conclude that if it is scattered, the source is
Galactic.  The source's spectrum and apparent diameter are shown in
Figure~\ref{fig:sourceJ}; its spectral index ($S \propto
\nu^{-\alpha}$) over the observed frequency range is $\alpha \approx
1$.  The spectrum shown in Figure~\ref{fig:sourceJ} is constructed
from observations with considerably different resolutions over a time
span of approximately 10~yrs.  Thus, it should be taken as indicative
of the actual spectrum, though a spectrum found from observations with
matched resolutions might not be quite as steep.

\subsubsection{Summary of Observations}

In addition to the observations summarized here, we have obtained
additional observations of this source at~6 and~20~cm (and~90~cm) with
the VLA with slightly higher resolutions than those described here.
These observations are in the process of being reduced and the results
will be presented elsewhere (\cite{lcka98}).  However, preliminary
indications are that source structure and diameters from these
observations are consistent with those reported here.

\paragraph{1.3cm}

In this paper we report a non-detection of this source in a VLBI
experiment.  This non-detection is consistent with the source's
spectrum extrapolated from longer wavelengths.  Our sensitivity was
50~mJy, and the source's flux is expected to be approximately 10~mJy.

\paragraph{2cm}

Yusef-Zadeh et al.~(1998) present a VLA image with a resolution of
$0\farcs36 \times 0\farcs13$.  The source is a double with separation
of approximately 4\arcsec.  The northern component, component~A, is
the brighter of the two and has a complex structure, approximately
0\farcs5 in diameter with a ``barrel-shaped appearance.''  The
southern component, component~B, is possibly resolved into a compact
core and a weak halo.

Anantharamaiah \& Goss~(1994, private communication) imaged the source
with the VLA, though with a resolution of only $2\farcs6 \times
1\farcs2$.  Component~A appears unresolved and has a flux of 9.9~mJy.
Component~B is slightly resolved and has a flux of approximately
2~mJy.

\paragraph{3.5cm}

Anantharamaiah \& Goss~(1994, private communication) also obtained VLA
observations at 3.5~cm contemporaneously with the 2~cm observations.
At their resolution of $4\farcs6 \times 2\farcs7$, the source is a
partially resolved double, again with a separation of approximately
4\arcsec.  Component~A is again unresolved and has a flux of
approximately 20~mJy.  Component~B is, at best, mildly resolved and
has a flux of approximately 4~mJy.  There is a faint source visible,
with a flux density of approximately 1~mJy and located approximately
7\arcsec\ to the southeast, which may be a third component of this
source.

We were unable to detect this source in a VLBI experiment.  However,
the expected rms visibility for our VLBA-VLBA baselines is 15~mJy, so
a detection on these baselines would be only marginal.  On the
VLBA-VLA baselines, the rms visibility is 3~mJy, so we would have
expected this source to be approximately a $5\sigma$ detection.  As we
discuss in \S\ref{sec:gc.sourceJ}, we attribute this non-detection to
be the result of source structure.

\paragraph{6cm}

Yusef-Zadeh et al.~(1998) present both total and polarized intensity
images obtained with the VLA with a resolution of $0\farcs98 \times
0\farcs34$.  In total intensity, the source can still be identified as
a double, but only component~A is present in the polarized image.
Component~A has a diameter of approximately 0\farcs5 while component~B
is unresolved.  The total integrated flux is 28~mJy and the polarized
flux is 2.4~mJy (Yusef-Zadeh~1994, private communication).

Becker et al.~(1994) detect this source in their 6~cm survey of the
Galactic plane.  They find a flux of 37.2~mJy.  From a
naturally-weighted image with a resolution of approximately 4\arcsec,
they fit a single gaussian to obtain a source diameter of 1\farcs3.

We observed this source as part of our VLA program.  We find it to be
a nearly resolved double with a separation of 3\farcs7.  The
integrated flux is 41.9~mJy.  Fitting two gaussians to a
naturally-weighted image, we find the northern component to have a
diameter $0\farcs71 \times 0\farcs47$.  Our subsequent, slightly
higher resolution observations indicate that component~A's diameter is
approximately 0\farcs4.

We did not detect this source in a VLBI experiment.  The expected rms
visibility for our VLBA-VLBA baselines is 10~mJy, so that on these
baselines we would expect a $3\sigma$ detection, and a $15\sigma$
detection on VLBA-VLA baselines.  The higher frequency observations
described above indicate that the probable cause for the non-detection
is source structure.

\paragraph{18--21cm}

At~1400~MHz the source has a flux of 267~mJy.  When fit with a single
gaussian, the source has a diameter of 3\farcs8; the resolution of
these observations was approximately 5\arcsec\ (\cite{zhbwp90}).
However, the source was 20\farcm7 from the phase center of their
observation, so the flux density measurement has a relatively large
uncertainty because of the large primary beam correction.

Anantharamaiah \& Goss~(1994, private communication) also imaged the
source at~1477~MHz, though not contemporaneously with the observations
described above.  With a resolution of approximately 5\arcsec, the
source is unresolved and has an integrated flux of 154~mJy.

We have observations of this source at~1281 and~1658~MHz from our VLA
program.  At~1281~MHz, the integrated flux is 66.9~mJy and,
at~1658~MHz, it is 83.1~mJy.  These are likely to be lower limits to
the actual values as inner $u$-$v$ cutoffs of 1--4~k$\lambda$ were
used in the imaging.  Super uniformly-weighted images at both 1281
and~1658~MHz, with resolutions of approximately $3\farcs2 \times
2\farcs6$, show the source to be unresolved.

\paragraph{35cm}

Gray~(1996, private communication) detected this source in the
843~MHz MOST survey (\cite{g94}).  The flux density of the source is
480~mJy.  The beam diameter of the MOST is too large, $90\arcsec \times
45\arcsec$, to derive a reliable angular diameter.

\paragraph{90cm}

Anantharamaiah et al.~(1991) used the the VLA to observe the
\hbox{GC}.  The VLA's primary beam at~90~cm is large enough that this
source was included in the field of view.  They find a flux of 580~mJy
for the source.  The diameter is $19\farcs8 \times 18\farcs6$
at~170\arcdeg\ (Anantharamaiah \& Goss~1994, private communication).
The beam is approximately 11\arcsec.  Based on their (brief)
description, this source is probably only mildly resolved, with a
``deconvolved'' diameter of 16\arcsec.

\paragraph{375cm}

The line of sight toward the source shows strong absorption and the
source cannot be identified (\cite{lk85}).

\subsubsection{Scattering}

The source is 15\arcmin\ from \sgra, with most of the displacement in
Galactic latitude, and is potentially seen through the scattering
region responsible for the enhanced broadening of \sgra\ and the OH/IR
stars (in particular, see Fig.~2 of \cite{fdcv94}).  The actual amount
of broadening expected for this source depends upon its distance from
the Sun and the Galactic center-scattering screen distance, $\delgc$.
Throughout we focus on component~A as it is stronger, and more
complete information exists for it.  The diameters summarized above
are shown in Figure~\ref{fig:sourceJ}.

We first determine if scattering dominates the observed angular
diameters.  We solve for the source's diameter dependence with
wavelength, $\theta \propto \lambda^\beta$.  If scattering determines
the observed diameter, we would expect $\beta = 2$.  Using the
measured diameters at~0.3 and 1.4~GHz, where scattering is most likely
to dominate, $\beta \approx 1.3$.  If we use this angular dependence
to predict the diameter at 6~cm, we find an angular diameter $\theta
\approx 0\farcs4$, comparable to that which is observed.  However,
this estimate does not allow for the intrinsic source structure
indicated by the higher frequency observations.

In order to allow for intrinsic source structure, we model the
apparent source diameter, $\theta_{\mathrm{app}}$, as a function of
frequency
\begin{equation}
[\theta_{\mathrm{app}}(\nu)]^2 
 = \left(\frac{\ths}{\nu_{\mathrm{GHz}}^2}\right)^2 + \thi^2
\label{eqn:size}
\end{equation}
with $\nu_{\mathrm{GHz}}$ the observation frequency in GHz, $\ths$ the
scattering angle at the nominal frequency of 1~GHz, and
$\thi$ the intrinsic diameter, i.e., the diameter the
source would have in the absence of any scattering, at 1~GHz.
Motivated by the complex structure seen at~2~cm (\cite{y-zcr98}), we
have assumed that the intrinsic source diameter is relatively
frequency independent.  Any intrinsic source structure whose frequency
dependence is of the form $\nu^\gamma$ with $\gamma < 0$ will only
strengthen our conclusions.  From the higher frequency observations
described above, we take $\thi = 0\farcs4$, the diameter of the
stronger component, component~A.

If the source is Galactic and the source is behind the scattering
screen, $\ths = 1\farcs3$, namely the diameter of \sgra.  In this case
we expect $\theta_{\mathrm{app}} = 0\farcs77$ at~1.4~GHz and
$\theta_{\mathrm{app}} = 12\arcsec$ at~0.33~GHz.  This predicted
1.4~GHz diameter is similar to that observed for \sgra.  More
importantly, the highest resolution observations at~1.4~GHz have a
resolution of~2\farcs9.  The beam is sufficiently larger than the
predicted diameter of the source that we expect the source to be
unresolved, as is indeed found.  The observed diameter at~0.33~GHz is
25\% larger than this value.  A larger diameter could be obtained if
the source-screen separation is slightly larger than that of the
\sgra-screen separation.  Cordes \& Lazio~(1997) show that a source a
distance~$\Delta$ behind the scattering screen has a scattering
diameter of $\ths \simeq \thgal(\Delta/\delgc)$.  If the
\sgra-scattering region separation is $\delgc = 150$~pc (\cite{lc98}),
then a source-scattering region separation of $\Delta = 190$~pc, i.e.,
the source would be 40~pc more distant than \sgra, would be sufficient
to explain the observed diameter at~0.33~GHz.

The scattering toward the GC is so intense that our analysis of the
0.33~GHz diameter does not depend upon what we assume for the
intrinsic diameter of the source.  Since the scattering diameter
determines the minimum diameter of the source in strong scattering
(\cite{cc74}), we could have taken the intrinsic diameter to be the
4\arcsec\ separation of the two components.  If we had done so, our
prediction for the apparent diameter at~0.33~GHz would have increased
to only $\theta_{\mathrm{app}} = 12\farcs5$, and our conclusions would
not have been affected.

If the source is extragalactic \emph{and} behind the \sgra\ scattering
screen, then, from equation~(\ref{eqn:xgalsize}), the minimum value of
$\ths$ is approximately double the diameter of \sgra\ or 2\farcs6,
which occurs when the scattering region is midway between the Sun and
the \hbox{GC}.  Even this minimal value of scattering for an
extragalactic source is too much, however.  In this case, the
predicted apparent diameter at~0.33~GHz is $\theta_{\mathrm{app}} =
24\arcsec$, 50\% larger than the diameter of 16\arcsec.

We conclude that this source is unlikely to be extragalactic
\emph{and} affected by the scattering screen in front of \sgra.  Its
morphology is suggestive of an extragalactic source, in which case it
would indicate that the extent of the scattering screen to positive
latitudes is no more than 15\arcmin\ (implying an axial ratio for the
scattering screen of at least 0.6) or that the scattering material is
patchy.  Alternately, the source could be Galactic, in which case it
is likely to be no more than approximately 40~pc more distant than
\sgra.

\subsection{1LC~358.600$-$0.060}

This source is within the star-formation region Sgr~E and appears to
be a central point source surrounded by a shell,
Figure~\ref{fig:1742-3009}.  Previous observations at 1.4~GHz could
not resolve the point source (\cite{gwcg93}).

On the basis of a nearly flat spectral index, Gray et al.~(1993)
conclude that this source is an \ion{H}{2} region.  They determine a
shell diameter of nearly 100\arcsec, and their image shows
considerable diffuse emission in the region of this source.  Our image
of the shell-like structure shows a more limited extent.  We have
formed a spectral-index map from our images at~1281 and~1658~MHz.
Much, but not all, of the shell is marked by a steep spectral index,
$\alpha \gtrsim 1$.  The central point source also appears to have a
steep spectral index, though it is not distinguished clearly from the
rest of the structure.

From the morphology and spectral information, we classify this object
tentatively as a supernova remnant.  Polarization information would be
useful in making a final determination.

\subsection{GPSR~358.960$+$0.555}

This object is a compact, 8--10~mJy source found by Zoonematkermani et
al.~(1990).  The source is located 4\arcmin\ from the phase center of
their field 359.0$+$0.5, so that the primary beam correction is not
large.  This position is 2\arcmin\ from the phase center of our field
358.9$+$0.5.  Within a 30\arcsec\ radius of the quoted position for
this source, we find no source brighter than 1.5~\mjybm.

Zoonematkermani et al.'s~(1990) minimum detectable flux for this field
is 10~mJy.  This source is either a misidentified noise fluctuation or
another Galactic radio transient (\cite{dwben76}; \cite{zetal92}).

\section{Conclusions}\label{sec:gc.conclude}

This paper has reported the results of a program to identify and
obtain scattering diameters for extragalactic sources seen through the
Galactic center scattering region.  Because they are located far
behind the GC, the scattering diameters of extragalactic sources, when
compared to the scattering diameter of GC sources such as \sgra, can
constrain the \emph{radial} location of the scattering region, viz.\
equation~(\ref{eqn:xgalsize}) and Figure~\ref{fig:xgalsize}.

Using the VLA we observed 10 (11) fields at~20~cm (6~cm) containing
15 suspected extragalactic sources.  We increased our catalog of sources
to well over 100 through the use of pdf\clean: The intensity histogram of
the primary beam was used to identify positive brightness image pixels
that produced deviations from the shape of the expected noise-only
histogram.  We found approximately 10 sources per field.

Follow-up VLBI observations on a subset of these sources have
determined the scattering diameters for two heavily scattered
extragalactic sources.  Their diameters are too small, by factors of
4--10, for them to be seen through the scattering region in front of
\sgra.  However, they can be used, in conjunction with the heavily
scattered masers, to set constraints on the angular extent of the
region.

Our fields show a paucity of sources near \sgra; a previous survey
with more uniform sky coverage, but at a lower sensitivity also shows
a paucity.  Such a deficit could arise if the scattering toward the GC
is so severe that our (and previous) observations resolve out
extragalactic sources.  The sources reported here are combined with
angular broadening measurements of \sgra\ and OH masers and free-free
emission and absorption measurements from the literature.  These data
are then used in a likelihood analysis to determine the model
parameters of the GC scattering region (\cite{lc98}).

\acknowledgements
We thank M.~Goss and F.~Yusef-Zadeh for contributing information on
1LC~359.872$+$0.178.  We thank the many GC observers who contributed
potential extragalactic sources during the early stages of this
project.  We thank N.~Bartel for contributing additional angular
broadening measurements.  H.~Liszt contributed data prior to
publication.  The Very Large Array (VLA) and the Very Long Baseline
Array (VLBA) are facilities of the National Science Foundation
operated under cooperative agreement by Associated Universities, Inc.
This research made use of the Simbad database, operated at the CDS,
Strasbourg, France.  This research was supported by the NSF under
grant AST~92-18075 and AST~95-28394.  The NAIC is operated by Cornell
University under a cooperative agreement with the \hbox{NSF}.
TJWL holds a National Research Council-NRL Research Associateship.
Basic research in astronomy at the Naval Research Laboratory is
supported by the Office of Naval Research.

\clearpage 

\begin{figure}
\caption[Extragalactic Source Diameters]
{The diameter of an extragalactic source at 1.4~GHz seen through the scattering
region in front of \sgra\ as a function of the Galactic
center-scattering region distance, $\delgc$.  The dotted line
indicates an extreme lower limit on $\delgc$ as derived from the lack
of free-free absorption toward \sgra\ at centimeter wavelengths.  At
this frequency, the scattering diameter of \sgra\ is 0\farcs7.  The
scattering diameter scales as $\ths \propto \nu^{-2}$.}
\label{fig:xgalsize}
\end{figure}

\begin{figure}
\caption[VLA Fields and VLBI Sources Observed]
{Fields observed with the VLA and sources observed with the \hbox{VLBA}.
Large, solid circles show the half-power primary VLA beam width at 1658~MHz.
Small, dashed circles show the half-power primary beam at 4863~MHz.
Crosses are the VLBI program sources and asterisks are the control sources
\sgra\ and B1741$-$312.  The contours are from the 5~GHz survey with
the NRAO 91~m telescope (Condon, Broderick, \& Seielstad~1991) and are
at~0.32, 0.64, 1.28, 2.56, 5.12, 10.2, 20.5, 41, and~82\% of the
peak, 250~Jy~beam${}^{-1}$.  Breaks in the contours result from gaps
in the original survey.
}
\label{fig:point}
\end{figure}

\begin{figure}
\caption[Sources Detected in Radio Surveys of the Galactic Center]
{Radio sources detected in this work and by Zoonematkermani
et al.~(1990) and Helfand et al.~(1992).  Sources are represented by
\textit{asterisks} if they are Galactic, \textit{circles} if
extragalactic, and by \textit{triangles} if unidentified.  See text
for criteria for distinguishing Galactic from extragalactic sources.
The large circles are the fields observed at~20~cm, cf.\
Figure~\ref{fig:point}.}
\label{fig:srccat}
\end{figure}

\begin{figure}
\caption[]{Correlated flux density as a function of projected baseline
for B1739$-$298 at 6~cm.  The solid line shows a circular gaussian
with a diameter of 27.9~mas.}
\label{fig:j1742-2949}
\end{figure}

\begin{figure}
\caption[]{1LC~359.872$+$0.178.  (\textit{a}) The spectrum of the
northern component. ({b}) The apparent diameter of the northern
component.  The dashed line shows the angular diameter dependence if
angular broadening dominates the source size.}
\label{fig:sourceJ}
\end{figure}

\begin{figure}
\caption[]{1LC~358.600$-$0.060, a shell-like source with an embedded
point source detected in our survey.
(\textit{Top}) Total intensity.  Contours are in units of 0.96~\mjybm
$\times$ $-2$, 2, 3, 5, 7.07, 10,~\ldots, the size of the beam is
shown in the lower left. 
(\textit{Bottom}) Spectral index between~1281 and~1658~MHz.  Contours
are $-1.5$, $-1.0$, $-0.5$, 0.5, 1.0, and~1.5; negative contours are
dashed.}
\label{fig:1742-3009}
\end{figure}

\clearpage

\begin{deluxetable}{lccc}
\tablecaption{Pointing Centers\label{tab:point}}
\tablehead{\colhead{Field} 
	& \colhead{RA} & \colhead{Dec} & \colhead{\sgra\ Distance}\\
	& \colhead{(J2000)} & \colhead{(J2000)} & \colhead{(\arcmin)}}
\startdata

 357.9$-$1.0                  & 17 44 23.5 & $-$31 16 35 & 137 \nl 
 358.1$-$0.0\tablenotemark{b} & 17 41 17.0 & $-$30 33 00 & 113 \nl 
 358.2$-$0.0\tablenotemark{a} & 17 41 23.0 & $-$30 30 02 & 110 \nl 
 358.7$-$0.0                  & 17 42 34.0 & $-$30 05 38 & \phn80 \nl 
 358.9$+$0.5                  & 17 40 54.0 & $-$29 29 49 & \phn77 \nl 

\nl
              	     	 
 359.9$+$0.2                  & 17 44 36.0 & $-$28 57 10 & \phn16 \nl 
 \phn\phn0.0$+$0.0\tablenotemark{a} & 17 45 30.0 & $-$28 54 39 & \phn\phn6 \nl 
 \phn\phn0.0$+$0.0\tablenotemark{b} & 17 45 36.0 & $-$28 53 20 & \phn\phn7 \nl 
 \phn\phn0.2$-$0.7            & 17 48 48.0 & $-$29 07 38 & \phn47 \nl 
 \phn\phn0.5$+$0.2            & 17 45 52.0 & $-$28 20 25 & \phn40 \nl 

\tableline
\tablebreak

 \phn\phn1.0$+$1.6            & 17 42 03.0 & $-$27 13 23 & 120 \nl 
 \phn\phn1.1$-$0.1\tablenotemark{a} & 17 48 33.0 & $-$28 02 30 & \phn72 \nl 
 \phn\phn1.2$-$0.0\tablenotemark{b} & 17 48 31.0 & $-$27 55 00 & \phn78 \nl 
 \phn\phn1.3$-$0.0\tablenotemark{a} & 17 48 51.0 & $-$27 51 55 & \phn84 \nl 
 \phn\phn3.7$+$0.6\tablenotemark{c} & 17 51 51.2 & $-$25 23 59 & 232 \nl 

\enddata

\tablenotetext{a}{Observed only at~6~cm.}
\tablenotetext{b}{Observed only at~20~cm.}
\tablenotetext{c}{Contains the phase calibrator, B1748$-$253.}
\end{deluxetable}

\begin{deluxetable}{lcccccccc}

\tablecaption{VLBI Sources\label{tab:vlbisources}}

\tablehead{ & \multicolumn{2}{c}{6~cm} 
	& \multicolumn{2}{c}{3.6~cm} 
	&\multicolumn{2}{c}{1.3~cm} \\
	\cline{2-3} \cline{4-5} \cline{6-7}
	\colhead{Name} 
	& \colhead{$S$} & \colhead{$\theta$}
	& \colhead{$S$} & \colhead{$\theta$}
	& \colhead{$S$} & \colhead{$\theta$} \\
	& \colhead{(mJy)} & \colhead{(mas)} 
	& \colhead{(mJy)} & \colhead{(mas)} 
	& \colhead{(mJy)} & \colhead{(mas)}  }

\startdata

1LC~358.439$-$0.211    	    & $<$\phn\phn5.9 	& $>$240\phd\phn    & $<$\phn\phn5.8       & $>$190\phd\phn	  & $<$\phn41 & $>$20\phd\phn \nl 
B1739$-$298\tablenotemark{a} & \phm{$<$}\phn55.6 & \phm{$<$}\phn27.9 & \phm{$<$}\phn\phn2.5 & $<$\phn55\phd\phn	  & $<$\phn54 & $>$60\phd\phn \nl
1LC~359.872$+$0.178   	    & $<$\phn\phn5.9 	& $>$240\phd\phn    & $<$\phn\phn3.5       & $>$190\phd\phn	  & $<$\phn32 & $>$58\phd\phn \nl 
1LC~0.846$+$1.173 	    & $<$\phn\phn6.9 	& $>$390\phd\phn    & $<$\phn16\phd\phn    & $>$160\phd\phn	  & $<$\phn32 & $>$60\phd\phn \nl 
	            		             			   
\nl
	       		       
B1741$-$312\tablenotemark{b} & \phm{$<$}406\phd\phn & \phm{$<$}\phn18.2 & \phm{$<$}530\phd\phn & \phm{$<$}\phn\phn7.2 & \phm{$<$}506 & \phn\phn1.3\nl
\sgra                       & \phm{$<$}420\phd\phn & \phm{$<$}\phn51.6 & \phm{$<$}510\phd\phn & \phm{$<$}\phn16.6    & \phm{$<$}860 & \phn\phn2.3\nl

\enddata

\tablenotetext{a}{Galactic coordinates $\ell = 358.918$, $b = 0.073$}
\tablenotetext{b}{Galactic coordinates $\ell = 357.862$, $b = -0.997$}

\end{deluxetable}

\begin{table}
\dummytable\label{tab:sources}
\end{table}

\begin{table}
\dummytable\label{tab:c-sources}
\end{table}

\end{document}


\setcounter{page}{27}
\setcounter{table}{2}

\begin{deluxetable}{lccccccccccl}

\ptlandscape
\footnotesize
\tablewidth{550pt}
\tablecaption{Sources from 1LC 20~cm Survey\label{tab:sources}}

\tablehead{ & &
	& \multicolumn{3}{c}{1658~MHz} & \multicolumn{3}{c}{1281~MHz} \\
	\cline{4-6} \cline{7-9}
	\colhead{Name} 
	& \colhead{RA} & \colhead{Dec}
	& \colhead{$I$} & \colhead{$S$} & \colhead{$\theta$}
	& \colhead{$I$} & \colhead{$S$} & \colhead{$\theta$}
	& \colhead{$\alpha$} & \colhead{$\Theta$} & \colhead{Notes} \\
	\colhead{(1)}
	& \colhead{(2)} & \colhead{(3)}
	& \colhead{(4)} & \colhead{(5)}
	& \colhead{(6)} & \colhead{(7)} & \colhead{(8)}
	& \colhead{(9)} & \colhead{(10)} & \colhead{(11)}
	& \colhead{(12)}}

\startdata

357.641$-$0.080 e &  17 40 12.132  &  $-$30 59 04.81  &         \nodata &         \nodata &         \nodata &    \phn\phn89.7 &          1279.3 &           118.9 &         \nodata &            30.7 \\
357.666$-$0.107 e &  17 40 22.284  &  $-$30 58 39.62  &         \nodata &         \nodata &         \nodata &    \phn\phn52.7 &          2783.9 &            60.7 &         \nodata &            29.1 & GPSR,Gal \\
357.695$-$0.084 e &  17 40 21.055  &  $-$30 56 27.60  &         \nodata &         \nodata &         \nodata &    \phn\phn52.9 &          1557.9 &            45.1 &         \nodata &            27.3 & GPSR,Gal \\
357.768$-$0.985 &  17 44 06.581  &  $-$31 21 12.52  & \phn\phn\phn7.9 & \phn\phn\phn8.0 &         \phn0.0 & \phn\phn\phn9.4 &    \phn\phn10.7 &         \phn3.3 &         \phs1.1 &         \phn6.3 \\
357.795$-$0.789 &  17 43 23.853  &  $-$31 13 42.71  &    \phn\phn21.0 &    \phn\phn46.3 &         \phn6.4 &    \phn\phn33.1 &    \phn\phn98.1 &            25.7 &         \phs2.9 &            15.2 & GPSR \\

\nl

357.806$-$1.047 &  17 44 26.965  &  $-$31 21 14.77  & \phn\phn\phn8.0 & \phn\phn\phn9.0 &         \phn0.0 & \phn\phn\phn7.7 & \phn\phn\phn6.3 &         \phn0.0 &      \phn$-$1.4 &         \phn4.7 & GPSR \\
357.808$-$0.299 &  17 41 29.028  &  $-$30 57 31.43  &    \phn\phn36.2 &    \phn\phn20.7 &         \phn0.0 &    \phn\phn53.4 &       \phn124.9 &         \phn9.7 &         \phs7.0 &            24.7 & GPSR \\
357.812$-$1.107 &  17 44 42.525  &  $-$31 22 48.60  &    \phn\phn17.7 &    \phn\phn19.9 &         \phn2.2 &    \phn\phn14.1 &    \phn\phn16.1 &         \phn2.6 &      \phn$-$0.8 &         \phn7.8 \\
\enddata

\tablecomments{(1) An ``e'' following a source's name indicates the
source is extended; ``m'' indicates that a source is one of multiple components.
(2) and (3) Right ascension and declination are epoch J2000.
(4) and (7) Peak intensities, $I$, are in units of \mjybm, with a
5\arcsec\ beam.
(5) and (8) Integrated fluxes, $S$, are in units of mJy.  
(6) and (9) Apparent sizes, $\theta$, are in units of arc$\,$seconds.
(10) Spectral index defined as $S \propto \nu^{-\alpha}$ and
calculated between 1281 and~1658~Mhz.  Extreme values probably reflect
the closely spaced observing frequences.
(11) Offset from phase center in arc$\,$minutes.
(12) Source cross-identifications---``GPSR'' indicates the source has
a counterpart in the Columbia Galactic plane survey, ``1LC'' indicates
a source also detected in our 6~cm survey, ``VLA'' indicates VLA
calibrators, ``Gal'' indicates sources likely to be Galactic, and
``xgal'' indicates sources likely to be extragalactic.}
\end{deluxetable}

\typeout{!Print this table in landscape mode!}